\documentclass[preprint,aps,12pt,showpacs,nofootinbib,tightenlines]{revtex4}
\usepackage{amsmath}
\usepackage{amssymb}
\usepackage{epsfig}
\usepackage{graphicx}
\begin{document}
\def\pslash{\rlap{\hspace{0.02cm}/}{p}}
\def\eslash{\rlap{\hspace{0.02cm}/}{e}}
\title{Direct $CP$ violation for $\bar{B}_{s}^{0}\rightarrow K^{0}\pi^{+}\pi^{-}$ decay in
QCD factorization
}
\thanks{ganglv@haut.edu.cn}%

\author{Gang L\"{u}}
\affiliation{%
 College of Science, Henan University of Technology, Zhengzhou 450001, China\\
 }%


\author{Bao-He Yuan}
\affiliation{
 North China University of Water Resources and Electric Power, zhengzhou 450011, China\\
}%

\author{Ke-Wei Wei}%
 \email{weikw@ihep.ac.cn}
\affiliation{
Institute of High Energy Physics, Chinese Academy of Sciences, Beijing 100049, China\\
}%



\date{\today}

\begin{abstract}
In the framework of QCD factorization, based on the first order of isospin violation,
 we study direct $CP$
violation in the decay of $\bar{B}_{s}^{0}
\rightarrow K^{0}\rho^{0}(\omega)\rightarrow K^{0}\pi^{+}\pi^{-}$
including the effect of
$\rho-\omega$ mixing. We find that the $CP$ violating asymmetry
is large via $\rho-\omega$ mixing mechanism
when the invariant mass of the $\pi^{+}\pi^{-}$ pair is in
the vicinity of the $\omega$ resonance. For the decay of $\bar{B}_{s}^{0}
\rightarrow K^{0}\rho^{0}(\omega)\rightarrow K^{0}\pi^{+}\pi^{-}$, the
maximum $CP$ violating asymmetries can reach about $46\%$.
We also  discuss the possibility  to observe the predicted
$CP$ violating asymmetries at the LHC.
\end{abstract}

\maketitle
\section{Introduction}
\label{intro}$CP$ violating asymmetry is one of the most important areas in
the decays of bottom hadrons. In the standard model(SM), a non-zero
complex phase in the Cabibbo-Kobayashi-Maskawa (CKM) matrix is
responsible for $CP$ violating phenomena. In recent years $CP$ violation
in several $B$ decays such as $B^0 \rightarrow J/\psi K_S^0$ and $B^0 \rightarrow
K^+ \pi^-$ has indeed been found in experiments
\cite{Belle,Yao2006}.  Due to its much higher statistics,
the Large Hadron Collider (LHC) will provide a new opportunity
to search for more $CP$ violation signals.

Direct $CP$ violating asymmetries in $b$-hadron decays occur through
the interference of at least two amplitudes with the weak phase
difference $\phi$ and the strong phase difference $\delta$. The weak
phase difference is determined by the CKM matrix while the strong
phase is usually difficult to control. In order to have a large $CP$
violating asymmetries signal, we have to apply  some phenomenological mechanism
to obtain a large $\delta$. It has been shown that the charge
symmetry violating mixing between $\rho^{0}$ and $\omega$ can be
used to obtain a large strong phase difference which is required for
large $CP$ violating asymmetries. Furthermore, it has been shown that the
measurement of the $CP$ violating asymmetries can be used to remove
the mod($\pi$) ambiguity in the determination of the $CP$ violating
phase angle $\alpha${\cite{Enomoto1996,Gardner1998,XH,Guo2000,Leitner2003}}.

Naive factorization approximation has been shown to be the leading order result
in the framework of QCD factorization when the radiative QCD
corrections of order $O(\alpha_{s}(m_b))$ $(m_b $ is the $b$-quark
mass) and the $O$($1/m_b$) corrections in the heavy quark effective
theory are neglected {\cite{qcdf}}. In naive factorization scheme,
 the hadronic matrix elements of
four-quark operators are assumed to be saturated by vacuum
intermediate states. Since the bottom hadrons are very heavy, their
hadronic decays are energetic. Hence the quark pair generated by one
current in the weak Hamiltonian moves very fast away from the weak
interaction point. Therefore, by the time this quark pair hadronizes
into a meson, it is already far away from other quarks and is
unlikely to interact with the remaining quarks. This quark pair is
factorized out and generates a meson {\cite{Bjorken1989,Dugan1991}}.
This approximation can only estimate the CP violation order neglecting
QCD correction. Furthermore, as pointed
out in previous studies {\cite{XH,Guo2000,Leitner2003}}, in order to
taken into account the nonfactorizable contributions, an effective
parameter, $N_c$, is introduced. The deviation of the value of $N_c$
from the color number, 3, measures the nonfactorizable effects in
the naive factorization scheme. Obviously, $N_c$ should depend on
the hadronization dynamics of different decay channels. In this scheme,
CP violation depends strongly on $N_c$ values, which makes the results uncertainties.

In the heavy quark limit, QCD factorization{\cite{qcdf}} includs nonfactorization strong
interaction correction, and the decay amplitudes can be calculated
at leading power in $\frac{\Lambda_{QCD}}{m_b}$ and at next-to-leading order in
$\alpha_{s}$, which can be expressed in terms of form factors and meson
light-cone distribution amplitudes. One can take
into account the nonfactorizable and chirally enhanced
hard-scattering spectator and annihilation contributions which
appear at order $O(\alpha_{s}(m_b))$ and  $O(1/m_b$), respectively.
In this work we adopt the QCD factorization
scheme including order-$\alpha_{s}$ correction to compute $CP$ violating asymmetry of
the decay $\bar{B}_{s}^{0}\rightarrow K^{0}\pi^{+}\pi^{-}$ via
the $\rho-\omega$ mixing mechanism. As will be shown
later, the $CP$ violating asymmetries in this decay channel could
be large and may be observed in the LHC
experiments.

The remainder of this paper is organized as follows. In Sec. II, we
present the form of the effective Hamiltonian and the general form of
QCD factorization. In Sec. III, we give the formalism for $CP$
violating asymmetries in $\bar{B}_{s}^{0}\rightarrow K^{0}\pi^{+}\pi^{-}$
decay. In Sec. IV, we calculate the branching ratio for decay process of
$\bar{B}_{s}^{0}\rightarrow K^{0}\rho^{0}(\omega)$ via $\rho-\omega$ mixing. We briefly
discuss the input parameters in Sec. V.
The numerical results are given in Sec. VI. In Sec. VII we discuss the
possibility to observe the predicted $CP$ violating asymmetries at
the LHC. Summary and conclusions are included in Sec. VIII.

\section{The effective Hamiltonian}
\label{sec:1} With the operator product expansion
{\cite{Buchalla1996}}, the effective Hamiltonian in bottom hadron
decays is
\begin{eqnarray}
 {\cal H}_{\rm eff}&=&\frac{G_{F}}{\sqrt{2}}
[\sum\limits_{p=u,c}\sum\limits_{q=d,s}V_{pb}V_{pq}^{*}(c_{1}O_{1}^{p}+c_{2}O_{2}^{p} \nonumber\\
&+& \sum\limits_{i=3}^{10}c_{i}O_{i}+c_{7\gamma}O_{7\gamma}+c_{8g}O_{8g}]
 +H_{.}c_{.},
\end{eqnarray}
where $c_{i}$ $(i=1,....,10,7\gamma,8g)$ are the Wilson coefficients, $V_{pb}$,
$V_{pq}$ are the CKM matrix elements. The
operators $O_{i}$ have the following form:
\begin{widetext}
\begin{equation}
\begin{array}{lll}
O_{1}^{p}=\bar{p}\gamma_{\mu}(1-\gamma_{5})b
\bar{q}\gamma^{\mu}(1-\gamma_{5})p,
&&O_{2}^{p}=\bar{p}_{\alpha}\gamma_{\mu}(1-\gamma_{5})b_{\beta}
\bar{q}_{\beta}\gamma^{\mu}(1-\gamma_{5})p_{\alpha}, \\
O_{3}=\bar{q}\gamma_{\mu}(1-\gamma_{5})b
\sum\limits_{q'}\bar{q'}\gamma^{\mu}(1-\gamma_{5})q',
&&O_{4}=\bar{q}_{\alpha}\gamma_{\mu}(1-\gamma_{5})b_{\beta}
\sum\limits_{q'}\bar{q'}_{\beta}\gamma^{\mu}(1-\gamma_{5})q'_{\alpha},\\
O_{5}=\bar{q}\gamma_{\mu}(1-\gamma_{5})b
\sum\limits_{q'}\bar{q'}\gamma^{\mu}(1+\gamma_{5})q',
&&O_{6}=\bar{q}_{\alpha}\gamma_{\mu}(1-\gamma_{5})b_{\beta}
\sum\limits_{q'}\bar{q'}_{\beta}\gamma^{\mu}(1+\gamma_{5})q'_{\alpha},\\
O_{7}=\frac{3}{2}\bar{q}\gamma_{\mu}(1-\gamma_{5})b
\sum\limits_{q'}e_{q'}\bar{q'}\gamma^{\mu}(1+\gamma_{5})q',
&&O_{8}=\frac{3}{2}\bar{q}_{\alpha}\gamma_{\mu}(1-\gamma_{5})b_{\beta}
\sum\limits_{q'}e_{q'}\bar{q'}_{\beta}\gamma^{\mu}(1+\gamma_{5})q'_{\alpha},\\
O_{9}=\frac{3}{2}\bar{q}\gamma_{\mu}(1-\gamma_{5})b
\sum\limits_{q'}e_{q'}\bar{q'}\gamma^{\mu}(1-\gamma_{5})q',
&&O_{10}=\frac{3}{2}\bar{q}_{\alpha}\gamma_{\mu}(1-\gamma_{5})b_{\beta}
\sum\limits_{q'}e_{q'}\bar{q'}_{\beta}\gamma^{\mu}(1-\gamma_{5})q'_{\alpha},  \\
O_{7\gamma}=\frac{-e}{8\pi^{2}}m_{b}\bar{s}\sigma_{\mu\nu}(1+\gamma_5)F^{\mu\nu}b,
&&O_{8g}=\frac{-g_s}{8\pi^{2}}m_{b}\bar{s}\sigma_{\mu\nu}(1+\gamma_5)G^{\mu\nu}b,
\end{array}
\end{equation}
\end{widetext}
where $\alpha$ and $\beta$ are color indices, $O_{1}^{p}$ and
$O_{2}^{p}$ are the tree operators, $O_{3}-O_{6}$ are QCD penguin
operators which are isosinglets, $O_{7}-O_{10}$ arise from
electroweak penguin operators which have both isospin $0$ and $1$
components. $O_{7\gamma}$ and $O_{8g}$ are the electromagnetic and chromomagnetic
dipole operators. $e_{q'}$ are the electric charges of the quarks and $q'=u,d,s,c,b$
is implied.

The Wilson coefficients can be calculated at a high scale $M_{W}$
and then evolved to scale $m_{b}$ using renormalization group equation.
In QCD factorization, We consider weak decay $B_{s}\to M_1 M_2$
($M_{1}$, $M_{2}$ refer to $K^{0}$ and $\rho^{0}$ mesons, respectively)
in the heavy-quark limit. Up to power corrections of order $\Lambda_{\rm QCD}/m_b$, the transition
matrix element of an operator ${\cal O}_i$ in the weak effective
Hamiltonian is given by{\cite{qcdf}}
\begin{eqnarray}
\label{fff}
\langle M_1 M_2|{\cal O}_i|\bar{B}\rangle &=&
\sum_j F_j^{B\to M_1}(m_2^2)\,\int_0^1 du\,T_{ij}^I(u)\,\Phi_{M_2}(u)\nonumber\\
\,\,&+&\,\,(M_1\leftrightarrow M_2)\nonumber\\
&&\hspace*{-2cm}
+\,\int_0^1 d\xi du dv \,T_i^{II}(\xi,u,v)\,
\Phi_B(\xi)\,\Phi_{M_1}(v)\,\Phi_{M_2}(u) \nonumber\\
&&\mbox{if $M_1$ and $M_2$ are both light,}
\end{eqnarray}
Here $F_j^{B\to M_{1,2}}(m_{2,1}^2)$ denotes a $B\to M_{1,2}$ form factor,
and $\Phi_X(u)$ is the light-cone distribution amplitude for the
quark-antiquark Fock state of meson $X$.
$T_{ij}^I(u)$ and $T_i^{II}(\xi,u,v)$ are hard-scattering functions,
which are perturbatively calculable. The hard-scattering kernels and
light-cone distribution amplitudes (LCDA) depend on a factorization scale
and scheme, which is suppressed in the notation of (\ref{fff}).
Finally, $m_{1,2}$ denote the light meson masses.

We match the effective weak Hamiltonian onto a transition operator, the matrix element is given by
($\lambda_p^{(D)}=V_{pb}V_{pD}^{*}$ with $D=d$ or $s$)
\begin{equation}\label{Top}
   \langle M_1'M_2'|{\cal H}_{\rm eff}|\bar B\rangle
   = \sum_{p=u,c} \lambda_p^{(D)}\,
   \langle M_1' M_2'|{\cal T}_A^p + {\cal T}_B^p|\bar B\rangle \,.
\end{equation}

Using the unitarity relation
\begin{eqnarray}
\lambda_u^{(D)}+\lambda_c^{(D)}+\lambda_t^{(D)}=0
\end{eqnarray}
we can get

\begin{eqnarray}\label{alphaidef4}
&&\sum_{p=u,c}\lambda_p^{(D)} {\cal T}_A^p
  =\sum_{p=u,c} \lambda_p^{(D)}\Bigg[\delta_{pu}\,\alpha_1(M_1 M_2)\,A([\bar q_s u][\bar u D])
    +\delta_{pu}\,\alpha_2(M_1 M_2)\,A([\bar q_s D][\bar u u])\Bigg]   \nonumber\\
     &+&  \lambda_u^{(D)} \Bigg[(\alpha_4^u(M_1 M_2)
     -\alpha_4^c(M_1 M_2))\,\sum_q A([\bar q_s q][\bar q D])
   +(\alpha_{4,\rm EW}^u(M_1 M_2)-\alpha_{4,\rm EW}^c(M_1 M_2))\,\sum_q\frac32\,e_q\,
    A([\bar q_s q][\bar q D]) \Bigg]  \nonumber\\
    &-& \lambda_t^{(D)}\Bigg[\alpha_3^c(M_1 M_2)\,\sum_q A([\bar q_s D][\bar q q])
    +\alpha_4^c(M_1 M_2)\,\sum_q A([\bar q_s q][\bar q D])
   +\alpha_{3,\rm EW}^c(M_1 M_2)\,\sum_q\frac32\,e_q\,
    A([\bar q_s D][\bar q q])   \nonumber\\
    &+& \alpha_{4,\rm EW}^c(M_1 M_2)\,\sum_q\frac32\,e_q\,
    A([\bar q_s q][\bar q D]) \Bigg]
\end{eqnarray}

where the sums extend over $q=u,d,s$, and $\bar q_s$ denotes the
spectator antiquark. The operators $A([\bar q_{M_1} q_{M_1}][\bar q_{M_2} q_{M_2}])$ also contain an
implicit sum over $q_s=u,d,s$ to cover all possible $B$-meson initial
states.

Next we need change the annihilation part

\begin{eqnarray}\label{bis}
   \sum_{p=u,c}\lambda_p^{(D)}{\cal T}_B^p
   &=& \sum_{p=u,c}\lambda_p^{(D)}    \nonumber\\
  &\times& \Bigg[\delta_{pu}\,b_1(M_1 M_2)\,\sum_{q'}
    B([\bar u q'][\bar q' u][\bar D b]) \nonumber\\
    &+& \delta_{pu}\,b_2(M_1 M_2)\,\sum_{q'}
    B([\bar u q'][\bar q' D][\bar u b]))\Bigg] \nonumber\\
  &-&\lambda_t^{(D)}\Bigg[ b_3(M_1 M_2)\,\sum_{q,q'} B([\bar q q'][\bar q' D][\bar q b])\nonumber\\
    &+& b_4(M_1 M_2)\,\sum_{q,q'} B([\bar q q'][\bar q' q][\bar D b])
    \nonumber\\
   &+& b_{3,\rm EW}(M_1 M_2)\,\sum_{q,q'} \frac{3}{2}\,e_q\,
    B([\bar q q'][\bar q' D][\bar q b])  \nonumber\\
    &+& b_{4,\rm EW}(M_1 M_2)\,\sum_{q,q'} \frac{3}{2}\,e_q\,
    B([\bar q q'][\bar q' q][\bar D b])\Bigg] \nonumber\\
\end{eqnarray}
where $b_{i}$, $b_{i,\rm EW}$ and $B$ are given by following.
The coefficients of the flavor operators $\alpha_i^p$ can be expressed in
terms of the coefficients $a_i^p$ defined in {\cite{qcdf}} as
follows:
\begin{eqnarray*}\label{ais}
   \alpha_1(M_1 M_2) &=& a_1(M_1 M_2) \,, \nonumber\\
   \alpha_2(M_1 M_2) &=& a_2(M_1 M_2) \,, \nonumber\\
   \alpha_3^p(M_1 M_2) &=& \left\{
    \begin{array}{cl}
     a_3^p(M_1 M_2) + a_5^p(M_1 M_2) \,;\\
       \quad \mbox{if~} M_1 M_2=PV  \,,
    \end{array}\right. \nonumber\\
   \alpha_4^p(M_1 M_2) &=& \left\{
    \begin{array}{cl}
     a_4^p(M_1 M_2) + r_{\chi}^{M_2}\,a_6^p(M_1 M_2) \,;\\
       \quad \mbox{if~} M_1 M_2=PV \,, \\
    \end{array}\right.\\
   \alpha_{3,\rm EW}^p(M_1 M_2) &=& \left\{
    \begin{array}{cl}
     a_9^p(M_1 M_2) + a_7^p(M_1 M_2) \,;\\
       \quad \mbox{if~} M_1 M_2=PV  \,,
    \end{array}\right. \nonumber\\
    \end{eqnarray*}
\begin{eqnarray}
   \alpha_{4,\rm EW}^p(M_1 M_2) &=& \left\{
    \begin{array}{cl}
     a_{10}^p(M_1 M_2) + r_{\chi}^{M_2}\,a_8^p(M_1 M_2) \,;\\
       \quad \mbox{if~} M_1 M_2=PV \,, \\
     \end{array}\right.
\end{eqnarray}
For pseudoscalar (P) meson $M_1$, the ratios
$r_\chi^{M_1}$ are defined as
\begin{equation}\label{rchi}
   r_\chi^{M_1}(\mu) = \frac{2m_{M_1}^2}{m_b(\mu)\,(m_q+m_s)(\mu)} \,,
\end{equation}
 All quark masses are running masses defined in the
$\overline{\rm MS}$ scheme, and $m_q$ denotes the average of the up and
down quark masses. For vector (V) meson $M_2$ we have
\begin{equation}
\label{rchiV}
   r_\chi^{M_2}(\mu) = \frac{2m_V}{m_b(\mu)}\,\frac{f_V^\perp(\mu)}{f_V} \,,
\end{equation}
where the scale-dependent transverse decay constant $f_V^\perp $ is defined as
\begin{equation}
   \langle V(p,\varepsilon^*)|\bar{q}\sigma_{\mu\nu} q'|0\rangle
   = f_V^\perp (p_\mu\varepsilon^*_\nu-p_\nu\varepsilon^*_\mu) \,.
\end{equation}
Note that all the terms proportional to $r_\chi^{M_2}$ are formally
suppressed by one power of $\Lambda_{\rm QCD}/m_b$ in the heavy-quark
limit.

The general form of the coefficients $a_i^p$ at next-to-leading order in
$\alpha_s$ is
\begin{eqnarray}
   a_i^p(M_1 M_2) &=& \left( C_i + \frac{C_{i\pm 1}}{N_c} \right)
   N_i(M_2) \nonumber\\
  && + \,\frac{C_{i\pm 1}}{N_c}\,\frac{C_F\alpha_s}{4\pi}
   \left[ V_i(M_2) + \frac{4\pi^2}{N_c}\,H_i(M_1 M_2) \right]\nonumber\\
   &&+ P_i^p(M_2) \,,
\end{eqnarray}
where $N_{c}$ is the number of colors, the upper (lower) signs
apply when $i$ is odd (even). It is
understood that the superscript `$p$' is to be omitted for $i=1,2$. The
quantities $V_i(M_2)$ account for one-loop vertex corrections,
$H_i(M_1 M_2)$ for hard spectator interactions, and $P_i^p(M_1 M_2)$ for
penguin contractions. The $ N_i(M_2)$ and $C_{F}$ are given by
\begin{equation}\label{loterms}
   N_i(M_2) = \Bigg\{
   \begin{array}{ll}
    ~0 \,; & \quad \mbox{$i=6,8 $ and $M_2=V$,} \\
    ~1 \,; & \quad \mbox{all other cases.}
   \end{array}
\end{equation}
\begin{equation}\label{CF}
  C_{F}=\frac{N_c^2-1}{2N_c}.
\end{equation}

The vertex corrections are given by{\cite{qcdf}}
\begin{equation}
V_i(M_2)=\left\{\begin{array}{ll}
     \int_0^1dx\,\Phi_{M_2}(x)\Big[12\ln\frac{m_b}{\mu} - 18 + g(x) \Big]
                 \\ (i=1-4,9,10),  \\
  \int_0^1dx\,\Phi_{M_2}(x)\Big[- 12\ln\frac{m_b}{\mu} + 6 - g(1-x) \Big]
       \\ (i=5,7),   \\
  \int_0^1dx\,\Phi_{m_2}(x)\Big[-6 + h(x) \Big]
       \\ (i=6,8),
       \end{array}\right.
\end{equation}
with
\begin{eqnarray}
   g(x)&=&3\Big( \frac{1-2x}{1-x}\ln x-i\pi \Big)
  + \Big[ 2 \,\mbox{Li}_2(x)
  - \ln^2\!x  \nonumber\\
    &+& \frac{2\ln x}{1-x} - (3+2i\pi)\ln x - (x\leftrightarrow 1-x)
    \Big] ,
\end{eqnarray}
 \begin{eqnarray}
   h(x)= 2 \,\mbox{Li}_2(x) - \ln^2\!x - (1+2\pi i)\,\ln x
    - (x\leftrightarrow 1-x). \nonumber \\
\end{eqnarray}
The constants $-18$, $6$, $-6$ are
scheme dependent and correspond to using the NDR scheme for $\gamma_5$.
The light-cone distribution amplitude (LCDA) $\Phi_{M_2}$ is the leading-twist
amplitude of $M_{2}$, whereas $\Phi_{m_2}$ is the twist-3 amplitude.
LCDA for pseudoscalar and vector mesons of
twist-2 are
 \begin{eqnarray}
 \Phi_{P}(x,\mu) &=& 6x(1-x)\left[1+\sum_{n=1}^\infty
 a_n^{P}(\mu)C_n^{3/2}(2x-1)\right], \nonumber \\
 \Phi^{V}_{\parallel}(x,\mu) &=& 6x(1-x)\left[1+\sum_{n=1}^\infty
 a_n^{V}(\mu)C_n^{3/2}(2x-1)\right],  \nonumber \\
 \Phi^{V}_{\perp}(x,\mu) &=& 6x(1-x)\left[1+\sum_{n=1}^\infty
 a_n^{\perp,V}(\mu)C_n^{3/2}(2x-1)\right],\nonumber \\
\end{eqnarray}
and twist-3 ones
 \begin{eqnarray}
&& \Phi_p(x)=1, \qquad \Phi_\sigma(x)=6x(1-x), \nonumber \\
&&  \Phi_v(x,\mu)=3\left[2x-1+\sum_{n=1}^\infty
 a_{n}^{\bot,V}(\mu)P_{n+1}(2x-1)\right], \nonumber \\
\end{eqnarray}
where $C_n(x)$ and $P_n(x)$ are the Gegenbauer and Legendre polynomials, respectively. $a_n(\mu)$ are
Gegenbauer moments that depend on the scale $\mu$.
$\Phi^{V}_{\perp}(x,\mu)$ and $ \Phi^{V}_{\parallel}(x,\mu)$ are the transverse and longitudinal quark
distributions of the polarized mesons.

At order $\alpha_s$ a correction from penguin contractions is present
only for $i=4,6$. For $i=4$ we obtain
\begin{eqnarray}\label{PK}
   P_4^p(M_2) &=& \frac{C_F\alpha_s}{4\pi N_c}\Bigg\{
    C_1 \left[ \frac{4}{3}\ln\frac{m_b}{\mu}
    + \frac{2}{3}-G_{M_2}(s_p) \right]   \nonumber\\
    &+& C_{3} \left[ \frac{8}{3}\ln\frac{m_b}{\mu} + \frac{4}{3}
    - G_{M_2}(0) - G_{M_2}(1) \right] \nonumber\\
   &+& (C_4+C_6)
    \Bigg[ \frac{4n_f}{3}\ln\frac{m_b}{\mu}
    - (n_f-2)G_{M_2}(0)   \nonumber\\
    &-& G_{M_2}(s_c) - G_{M_2}(1) \Bigg]\nonumber\\
    &-& 2 C_{8g}^{\rm eff} \int_0^1 \frac{dx}{1-x}\,
    \Phi_{M_2}(x) \Bigg\} \,,
\end{eqnarray}
where $n_f=5$ is the number of light quark flavors, and $s_u=0$,
$s_c=(m_c/m_b)^2$ are mass ratios involved in the evaluation of the
penguin diagrams. The function
$G_{M_2}(s)$ is given by
\begin{eqnarray}\label{GK}
   G_{M_2}(s)&=& \int_0^1\!dx\,G(s-i\epsilon,1-x)\,\Phi_{M_2}(x) \,, \\
   G(s,x)&=& -4\int_0^1\!du\,u(1-u) \ln[s-u(1-u)x] \nonumber\\
   &=& \frac{2(12s+5x-3x\ln s)}{9x}  \nonumber\\
   & -& \frac{4\sqrt{4s-x}\,(2s+x)}{3x^{3/2}}
    \arctan\sqrt{\frac{x}{4s-x}} \,.
\end{eqnarray}
For $i=6$, if $M_2$ is a vector meson, the result for the penguin contribution is
\begin{eqnarray}
   P_6^p(M_2) &=& - \frac{C_F\alpha_s}{4\pi N_c}\,\Bigg\{
    C_1\,\hat G_{M_2}(s_p)
    + C_3\,\Big[ \hat G_{M_2}(0) + \hat G_{M_2}(1) \Big] \nonumber\\
   &&\mbox{}+ (C_4+C_6) \Bigg[ (n_f-2)\,\hat G_{M_2}(0)
    + \hat G_{M_2}(s_c) \nonumber\\
    &+& \hat G_{M_2}(1) \Bigg] \Bigg\}.
\end{eqnarray}
 In analogy with (\ref{GK}), the function
$\hat G_{M_2}(s)$ is defined as
\begin{equation}
   \hat G_{M_2}(s) = \int_0^1\!dx\,G(s-i\epsilon,1-x)\,\Phi_{m_2}(x) \,.
\end{equation}

Electromagnetic corrections
 are present for $i=8,10$ and
correspond to the penguin diagrams. For
$i=10$ we obtain
\begin{eqnarray}\label{PKEW}
   P_{10}^p(M_2)& =& \frac{\alpha}{9\pi N_c}\,\Bigg\{
   (C_1+N_c C_2) \Bigg[ \frac{4}{3}\ln\frac{m_b}{\mu} \nonumber \\
   &+& \frac23
   - G_{M_2}(s_p) \Bigg]
    - 3 C_{7\gamma}^{\rm eff} \int_0^1 \frac{dx}{1-x}\,\Phi_{M_2}(x)
  \Bigg\} .\nonumber \\
\end{eqnarray}
For $i=8$
\begin{equation}
   P_8^p(M_2) = - \frac{\alpha}{9\pi N_c}\,(C_1+N_c C_2)\,
   \hat G_{M_2}(s_p),
\end{equation}
if $M_2$ is a vector meson.

The correction from hard gluon exchange between $M_2$ and the spectator
quark is given by
\begin{eqnarray}\label{hardspecterms1}
   H_i(M_1M_2)
  &=&\frac{B_{M_1 M_2}}{A_{M_1 M_2}}\,\frac{m_B}{\lambda_B}\,
   \int_0^1\!dx \int_0^1\!dy \Bigg[
   \frac{\Phi_{M_2}(x)\Phi_{M_1}(y)}{\bar x\bar y}     \nonumber\\
   &+& r_\chi^{M_1}\,\frac{\Phi_{M_2}(x)\Phi_{m_1}(y)}{x\bar y} \Bigg],
\end{eqnarray}
for $i=1$--4,9,10.
\begin{eqnarray}\label{hardspecterms2}
   H_i(M_1M_2)
   &=& - \frac{B_{M_1 M_2}}{A_{M_1 M_2}}\,\frac{m_B}{\lambda_B}\,
   \int_0^1\!dx \int_0^1\!dy \Bigg[
   \frac{\Phi_{M_2}(x)\Phi_{M_1}(y)}{x\bar y} \nonumber\\
   &+& r_\chi^{M_1}\,\frac{\Phi_{M_2}(x)\Phi_{m_1}(y)}{\bar x\bar y}
   \Bigg],
\end{eqnarray}
for $i=5,7$, and $H_i(M_1M_2)=0$ for $i=6,8$. \\
where
$\lambda_B$ is defined by
\begin{equation}
   \int_0^1 \frac{d\xi}{\xi}\,\Phi_B(\xi)\equiv \frac{m_B}{\lambda_B}
\end{equation}
with $\Phi_B(\xi)$ is one of the two light-cone distribution amplitudes of
the $B$ meson.

If $M_1=P$, $M_2=V$, $f$ refers to decay constant of
 relevant meson, $A_{M_1 M_2}$ and $B_{M_1M_2}$ are given by
\begin{equation}\label{am1m2}
   A_{M_1 M_2} = i\,\frac{G_F}{\sqrt2}
    (-2)m_{M_{1}}\,\epsilon_{M_{1}}^*\!\cdot p_B\,F_{0}^{B\to M_1}(0) f_{M_2},
\end{equation}
\begin{eqnarray}
B_{M_1M_2}=-\frac{G_F}{\sqrt{2}}f_{B_s}f_{M_1}f_{M_2}.
\end{eqnarray}
where $m_{M_{1}}$ and $\epsilon_{M_{1}}$ are the mass and polarization vector of the vector meson.
$F_{0}^{B\to M_1}$ is the form factor for $B \rightarrow M_{1}$ transition.

 We recall that the term involving $r_\chi^{M_1}$ is
suppressed by a factor of $\Lambda_{\rm QCD}/m_b$ in heavy-quark power
counting. Since the twist-3 distribution amplitude $\Phi_{m_1}(y)$ does
not vanish at $y=1$, the power-suppressed term is divergent. We extract
this divergence by defining a parameter $X_H^{M_1}$ through
\begin{eqnarray}\label{XHdef}
   \int_0^1\frac{d y}{\bar y}\,\Phi_{m_1}(y)
   &=& \Phi_{m_1}(1)\,\int_0^1\frac{d y}{\bar y}  \nonumber \\
    &+& \int_0^1\frac{d y}{\bar y}\,\Big[ \Phi_{m_1}(y)-\Phi_{m_1}(1)
    \Big] \nonumber\\
   &\equiv& \Phi_{m_1}(1)\,X_H^{M_1}
    + \int_0^1\frac{d y}{[\bar y]_+}\,\Phi_{m_1}(y) \, . \nonumber \\
\end{eqnarray}
The remaining integral is finite (it vanishes for pseudoscalar mesons
since $\Phi_p(y)=1$), but $X_H^{M_1}$ is an unknown parameter
representing a soft-gluon interaction with the spectator quark.
Since $X_H^{M_1}$ varies within a certain
range (specified later) and $X_H^M\sim\ln(m_b/\Lambda_{\rm QCD})${\cite{qcdf}},
we treat the resulting variation of the
coefficients $\alpha_i^p$ as an uncertainty. We also assume that
$X_H^{M_1}$ is universal, i.e., that it does not depend on $M_1$ and on
the index $i$ of $H_i(M_1 M_2)$. For the convolution integrals, one can find
the results in Ref. {\cite{qcdf}}.

For the annihilation contribution, one can get{\cite{qcdf}}:
\begin{eqnarray}
b_3^p = \frac{C_F}{N_c^2} \Big[ C_3 A_1^i + C_5 (A_3^i+A_3^f)
    + N_c C_6 A_3^f \Big],
\end{eqnarray}
\begin{eqnarray}
b_{3,\rm EW}^p &= \frac{C_F}{N_c^2} \Big[ C_9 A_1^i
    + C_7 (A_3^i+A_3^f) + N_c C_8 A_3^f \Big].
\end{eqnarray}

The weak annihilation kernels exhibit endpoint divergences, which we
treat in the same manner as the power corrections to the hard spectator
scattering. The divergent subtractions are interpreted as
\begin{equation}\label{XAdef}
   \int_0^1 \frac{dy}{y}\to X_A^{M_1} \,, \qquad
   \int_0^1\!dy\,\frac{\ln y}{y}\to -\frac{1}{2}\,(X_A^{M_1})^2 \,,
\end{equation}
and similarly for $M_2$ with $y\to \bar x$. The treatment of weak
annihilation is model-dependent in the QCD factorization approach. We treat $X_A^M$
as an unknown complex number of order $\ln(m_b/\Lambda_{\rm QCD})$ and
make the simplifying assumption that this number is independent of the
identity of the meson $M_1$ and the weak decay vertex. Here,
\begin{eqnarray}
A_1^i &\approx& -A_2^i \approx 6\pi\alpha_s \bigg[
3\bigg( X_A - 4 + \frac{\pi^2}{3} \bigg)  \nonumber \\
&+& r_\chi^{M_1} r_\chi^{M_2} (X_A^2-2 X_A) \bigg] ,
\end{eqnarray}
\begin{eqnarray}
A_3^i &\approx& 6\pi\alpha_s \bigg[ -3 r_\chi^{M_2}
\bigg( X_A^2 - 2 X_A - \frac{\pi^2}{3} \nonumber \\
 &+& 4 \bigg)+r_\chi^{M_1} \bigg( X_A^2 - 2 X_A + \frac{\pi^2}{3} \bigg)
    \bigg] ,
\end{eqnarray}
 \begin{eqnarray}
A_3^f &\approx& -6\pi\alpha_s \bigg[3 r_\chi^{M_2}
(2 X_A-1) (2-X_A)\  \ \, \nonumber \\
&-& r_\chi^{M_1}\,(2 X_A^2 - X_A) \bigg]
\end{eqnarray}
and $A_1^f=A_2^f=0$. Here, $M_1$ is  $K^0$ meson and $M_2$ is
$\rho^0$ meson.

\section{$CP$ violation in $\bar{B}_{s}^{0}
\rightarrow K^{0}\pi^{+}\pi^{-}$ decay}
\label{sec:2}
\subsection{Formalism}
\label{sec:3}In the vector meson dominance model
{\cite{Sakurai1969}}, the photon propagator is dressed by coupling
to vector mesons. Based on the same mechanism, $\rho-\omega$ mixing
was proposed {\cite{Connell1997}}. The formalism for $CP$ violation
in the decay of a bottom hadron, $B_{s}$, will be reviewed in the
following. The amplitude for $B_{s}\rightarrow K^{0}\pi^{+}\pi^{-}$,
$A$, can be written as
\begin{eqnarray}
A=\langle\pi^{+}\pi^{-}K^{0}|H^{T}|\bar{B}_{s}\rangle+\langle\pi^{+}\pi^{-}K^{0}|H^{P}|\bar{B}_{s}\rangle,
\end{eqnarray}
where $H^{T}$ and $H^{P}$ are the Hamiltonians for the tree and
penguin operators, respectively. We define the relative magnitude
and phases between these two contributions as follows:
\begin{eqnarray}
A=\langle\pi^{+}\pi^{-}K^{0}|H^{T}|\bar{B}_{s}\rangle[1+re^{i\delta}e^{i\phi}],
\end{eqnarray}
where $\delta$ and $\phi$ are strong and weak phase differences,
respectively. The weak phase difference $\phi$ arises from the
appropriate combination of the CKM matrix elements:
$\phi=\arg[(V_{tb}V_{ts}^{*})/(V_{ub}V_{us}^{*})]$. The
parameter $r$ is the absolute value of the ratio of tree and penguin
amplitudes,
\begin{eqnarray}
r=\left|\frac{\langle\pi^{+}\pi^{-}K^{0}|H^{P}|\bar{B}_{s}\rangle}{\langle\pi^{+}\pi^{-}K^{0}|H^{T}|\bar{B}_{s}\rangle}\right|.
\end{eqnarray}
The amplitude for $B_s\rightarrow\bar{K^0}\pi^{+}\pi^{-}$ is
\begin{eqnarray}
\bar{A}=\langle\pi^{+}\pi^{-}\bar{K^{0}}|H^{T}|B_{s}\rangle+\langle\pi^{+}\pi^{-}\bar{K^0}|H^{P}|B_{s}\rangle.
\end{eqnarray}
Then, the CP violating asymmetry, $a$, can be written as
\begin{eqnarray}
a=\frac{|A|^{2}-|\bar{A}|^{2}}{|A|^{2}+|\bar{A}|^{2}}=\frac{-2r\sin\delta\sin\phi}{1+2r\cos\delta\cos\phi+r^{2}}.
\end{eqnarray}
We can see explicitly from Eq. (40) that both weak and strong phase
differences are needed to produce $CP$ violation. $\rho-\omega$
mixing has the dual advantages that the strong phase difference is
large and well known {\cite{Enomoto1996,Gardner1998}}. In this
scenario one has
\begin{eqnarray}
\langle\pi^{+}\pi^{-}K^{0}|H^{T}|\bar{B}_{s}\rangle=\frac{g_{\rho}}{s_{\rho}s_{\omega}}\tilde{\Pi}_{\rho\omega}(t_{\omega}
+t_{\omega}^{a})+\frac{g_{\rho}}{s_{\rho}}(t_{\rho}+t_{\rho}^{a}),\nonumber\\
\end{eqnarray}
\begin{eqnarray}
\langle\pi^{+}\pi^{-}K^{0}|H^{P}|\bar{B}_{s}\rangle=\frac{g_{\rho}}{s_{\rho}s_{\omega}}\tilde{\Pi}_{\rho\omega}(p_{\omega}
+p_{\omega}^{a})+\frac{g_{\rho}}{s_{\rho}}(p_{\rho}+p_{\rho}^{a}),\nonumber\\
\end{eqnarray}
where $t_{V}(V=\rho$ or $ \omega)$ is the tree amplitude and $p_{V}$
is the penguin amplitude for producing a vector meson, $V$.
$t_{V}^{a}(V=\rho$ or $ \omega)$ is the tree annihilation amplitude and $p_{V}^{a}$
is the penguin annihilation amplitude. $g_{\rho}$ is the coupling for $\rho^{0}\rightarrow \pi^{+}\pi^{-}$,
$\tilde{\Pi}_{\rho\omega}$ is the effective $\rho-\omega$ mixing
amplitude, and $s_{V} $ is from the inverse propagator of the vector
meson V,
\begin{eqnarray}
s_{V}=s-m_{V}^{2}+i m_{V}\Gamma_{V},
\end{eqnarray}
with $\sqrt{s}$ being the invariant mass of the $\pi^{+}\pi^{-}$
pair.

The direct $\omega \rightarrow \pi^{+}\pi^{-}$ is effectively
absorbed into $\tilde{\Pi}_{\rho\omega}$, leading to the explicit
$s$ dependence of $\tilde{\Pi}_{\rho\omega}$ {\cite{Maltman1996}}.
Making the expansion
$\tilde{\Pi}_{\rho\omega}(s)=\tilde{\Pi}_{\rho\omega}(m_{\omega}^{2})
+(s-m_{\omega}^{2}) \tilde{\Pi}'_{\rho\omega}(m_{\omega}^{2})$, the
$\rho-\omega$ mixing parameters were determined in the fit of
Gardner and O'Connell {\cite{S.Gardner1998}}: $\rm
Re\tilde{\Pi}_{\rho\omega}(m_{\omega}^{2})=-3500\pm300$ MeV$^{2}$,
$\rm Im\tilde{\Pi}_{\rho\omega}(m_{\omega}^{2})=-300\pm300$
MeV$^{2}$, and $\tilde{\Pi}'_{\rho\omega}(m_{\omega}^{2})=0.03\pm
0.04$. In practice, the effect of the derivative term is negligible.
From Eqs. (40)(41)(44)(45)(46) one has
\begin{eqnarray}
re^{i\delta}e^{i\phi}=\frac{\tilde{\Pi}_{\rho\omega}(p_{\omega}+p_{\omega}^{a})+s_{\omega}(p_{\rho}+p_{\rho}^{a})}
{\tilde{\Pi}_{\rho\omega}(t_{\omega}+t_{\omega}^{a})+s_{\omega}(t_{\rho}+t_{\rho}^{a})}.
\end{eqnarray}
Defining
\begin{equation}
\frac{p_{\omega}+p_{\omega}^{a}}{t_{\rho}+t_{\rho}^{a}}=r'e^{i(\delta_{q}+\phi)}, \;
\frac{t_{\omega}+t_{\omega}^{a}}{t_{\rho}+t_{\rho}^{a}}=\alpha e^{i\delta_{\alpha}}, \;
\frac{p_{\rho}+p_{\rho}^{a}}{p_{\omega}+p_{\omega}^{a}}=\beta e^{i\delta_{\beta}},
\end{equation}
where $\delta_{\alpha}$, $\delta_{\beta}$, and $\delta_{q}$ are
strong phases, one finds the following expression from Eq. (47):
\begin{eqnarray}
re^{i\delta}=r'e^{i\delta_{q}}\frac{\tilde{\Pi}_{\rho\omega}+\beta
e^{i\delta^{\beta}}s_{\omega}
}{s_{\omega}+\tilde{\Pi}_{\rho\omega}\alpha e^{i\delta_{\alpha}} }.
\end{eqnarray}
$\alpha e^{i\delta_{\alpha}}$, $\beta e^{i\delta_{\beta}}$, and
$re^{i\delta}$ will be calculated in the QCD factorization approach
later. With Eq. (49), we can obtain $r\sin\delta$ and $r\cos\delta$.
In order to get the $CP$ violating asymmetry, $a$, in Eq. (43),
$\sin\phi$ and $\cos\phi$ are needed. $\phi$ is determined by the
CKM matrix elements. In the Wolfenstein parametrization
{\cite{Wolfenstein1983}}, one has
\begin{eqnarray}
\sin\phi=\frac{\eta}{\sqrt{[\rho(1-\rho)-\eta^{2}]^{2}+\eta^{2}}},\\
\cos\phi=\frac{\rho(1-\rho)-\eta^{2}}{\sqrt{[\rho(1-\rho)-\eta^{2}]^{2}+\eta^{2}}}.
\end{eqnarray}.
\subsection{$CP$ violation via $\rho-\omega$ mixing}
\label{sec:4}In the following we will study the $CP$
violating asymmetries in the following decay:
$\bar{B}_{s}^{0}\rightarrow K^{0}\rho^{0}(\omega)\rightarrow
K^{0}\pi^{+}\pi^{-}$. With the Eq. (4)(6)(7)(8), we can
calculate the decay amplitudes in QCD factorization scheme.
The expressions for the $\bar{B}_{s}^{0}\rightarrow K^{0}\rho^{0}(\omega)$
amplitudes are given by
\begin{eqnarray}
\sqrt{2}A_{\bar{B}_{s}^{0}\rightarrow K^{0}\rho^{0}}&=&
A_{K^{0}\rho^{0}}(\delta_{pu}\alpha_2-\alpha_4^{p}+\frac{3}{2}
\alpha_{3,EW}^{p}\nonumber \\
&+&\frac{1}{2}\alpha_{4,EW}^{p}-\beta_3^{p}+\frac{1}{2}\beta_{3,EW}^{p}),
\end{eqnarray}
\begin{eqnarray}
\sqrt{2}A_{\bar{B}_{s}^{0}\rightarrow K^{0}\omega}&=&
A_{K^{0}\omega}(\delta_{pu}\alpha_2+2\alpha_3^{p}+\alpha_4^{p}+\frac{1}{2}
\alpha_{3,EW}^{p}\nonumber \\
&-&\frac{1}{2}\alpha_{4,EW}^{p}+\beta_3^{p}-\frac{1}{2}\beta_{3,EW}^{p}),
\end{eqnarray}
where
\begin{eqnarray}
A_{K^{0}\rho^{0}}=(-2)i\frac{G_{F}}{\sqrt{2}}m_{\rho^{0}}
\varepsilon^{*}_{\rho^{0}}\cdot p_{B}F_{0}^{B_s\rightarrow K^{0}}(0)f_{\rho^{0}},
\end{eqnarray}
\begin{eqnarray}
A_{K^{0}\omega}=(-2)i\frac{G_{F}}{\sqrt{2}}m_{\omega}
\varepsilon^{*}_{\omega}\cdot p_{B}F_{0}^{B_s\rightarrow K^{0}}(0)f_{\omega}.
\end{eqnarray}
Here $F_{0}$ denote $B_{s}\rightarrow K^{0}$ meson form factor. $m_{\rho^{0}}$, $m_{\omega}$
are the mass of $\rho^{0}$ and $\omega$ mesons. $\varepsilon^{*}_{\rho^{0}}$,
$\varepsilon^{*}_{\omega}$ correspond to polarizing vectors. $f$ refers to the decay
constant.
Then we can get
\begin{eqnarray}
\sqrt{2}A_{\bar{B}_{s}\rightarrow K^{0}\rho^{0}}&=&A_{K^{0}\rho^{0}}
\Bigg[\delta_{pu}a_{2,K^{0}\rho^{0}}-a_{4,K^{0}\rho^{0}}^p
-\gamma_{\chi}^{k^0}a_{6,K^{0}\rho^{0}}^p \nonumber\\
&+&\frac{3}{2}(a_{9,K^{0}\rho^{0}}^p
+a_{7,K^{0}\rho^{0}}^p)
+\frac{1}{2}(a_{10,K^{0}\rho^{0}}\nonumber\\
&+&\gamma_{\chi}^{k^0}
a_{8,K^{0}\rho^{0}}^p)-\beta_3^p+\frac{1}{2}\beta_{3,EW}^p\Bigg],
\end{eqnarray}
\begin{eqnarray}
\sqrt{2}A_{\bar{B}_{s}\rightarrow K^{0}\omega}&=&A_{K^{0}\omega}
\Bigg[\delta_{pu}a_{2,K^{0}\omega}+2(a_{3,K^{0}\omega}^p+a_{5,K^{0}\omega}^p)\nonumber\\
&+&a_{4,K^{0}\omega}^p+\gamma_{\chi}^{\omega}a_{6,K^{0}\omega}^p
+\frac{1}{2}(a_{7,K^{0}\omega}+a_{9,K^{0}\omega}) \nonumber\\
&-&\frac{1}{2}(a_{10,K^{0}\omega}
+\gamma_{\chi}^{\omega}a_{8,K^{0}\omega}^p)+\beta_3^p-\frac{1}{2}\beta_{3,EW}^p\Bigg]. \nonumber\\
\end{eqnarray}
where the form of the coefficients $a_i^p$ at next-to-leading order in
$\alpha_s$ is given by Eq.(12), which $M_{1}$ is $K^{0}$ meson and $M_{2}$ is $\rho^{0}$ meson.
$\beta_{i}$ is the weak annihilation contribution in QCD factorization. $\gamma_{\chi}$
is chirally-enhanced terms which we have denoted above.

From Eq. (6)(7)(48), one can get
\begin{eqnarray}
\alpha e^{i\delta_{\alpha}}=\frac{t_\omega+t_\omega^a}{t_\rho+t_\rho^a}
=\frac{Q_1}{Q_2}.
\end{eqnarray}

\begin{eqnarray}
Q_1&=&A_{K^{0}\omega}
\Bigg\{\delta_{pu}a_{2,K^{0}\omega}
+2(a_{3,K^{0}\omega}^u-a_{3,K^{0}\omega}^c     \nonumber\\
&+&a_{5,K^{0}\omega}^u-a_{5,K^{0}\omega}^c)
+(a_{4,K^{0}\omega}^u-a_{4,K^{0}\omega}^c)       \nonumber\\
&+&\gamma_{\chi}^{\omega}(a_{6,K^{0}\omega}^u-a_{6,K^{0}\omega}^c)
+\frac{1}{2}(a_{7,K^{0}\omega}^u-a_{7,K^{0}\omega}^c     \nonumber\\
&+&a_{9,K^{0}\omega}^u-a_{9,K^{0}\omega}^c)
-\frac{1}{2}\Bigg[a_{10,K^{0}\omega}^u-a_{10,K^{0}\omega}^c    \nonumber\\
&+&\gamma_{\chi}^{\omega}(a_{8,K^{0}\omega}^u-a_{8,K^{0}\omega}^c)\Bigg]\Bigg\}
\end{eqnarray}

\begin{eqnarray}
Q_2&=&A_{K^{0}\rho^{0}}
\Bigg\{\delta_{pu}a_{2,K^{0}\rho^{0}}
-(a_{4,K^{0}\rho^{0}}^u - a_{4,K^{0}\rho^{0}}^c)    \nonumber\\
&-&\gamma_{\chi}^{k^0}(a_{6,K^{0}\rho^{0}}^u-a_{6,K^{0}\rho^{0}}^c )
+\frac{3}{2}(a_{9,K^{0}\rho^{0}}^u-a_{9,K^{0}\rho^{0}}^c    \nonumber\\
&+&a_{7,K^{0}\rho^{0}}^u-a_{7,K^{0}\rho^{0}}^c)
+\frac{1}{2}\Bigg[a_{10,K^{0}\rho^{0}}^u-a_{10,K^{0}\rho^{0}}^c           \nonumber\\
&+&\gamma_{\chi}^{k^0}
(a_{8,K^{0}\rho^{0}}^u-a_{8,K^{0}\rho^{0}}^c)\Bigg]\Bigg\},
\end{eqnarray}

In a similar way, with the aid of the Fierz identities, we can
evaluate the penguin operator contributions $p_{\rho}$ and
$p_{\omega}$. From Eq. (48) we have
\begin{eqnarray}
\beta e^{i\delta_{\beta}}=\frac{p_{\rho}+p_{\rho}^{a}}{p_{\omega}+p_{\omega}^{a}}
=\frac{Q_3}{Q_4},
\end{eqnarray}
where
\begin{eqnarray}
Q_3&=&A_{K^{0}\rho^{0}}\Bigg[-a_{4,K^{0}\rho^{0}}^c
-\gamma_x^{k^0}a_{6,K^{0}\rho^{0}}^c \nonumber\\
&+&\frac{3}{2}(a_{9,K^{0}\rho^{0}}^c
+a_{7,K^{0}\rho^{0}}^c)
+\frac{1}{2}(a_{10,K^{0}\rho^{0}}^c\nonumber\\
&+&\gamma_x^{k^0}
a_{8,K^{0}\rho^{0}}^c)-\beta_3+\frac{1}{2}\beta_{3,EW}\Bigg],
\end{eqnarray}
\begin{eqnarray}
Q_4&=&A_{K^{0}\omega}\Bigg[2(a_{3,K^{0}\omega}^c+a_{5,K^{0}\omega}^c)\nonumber\\
&+&a_{4,K^{0}\omega}^c+\gamma_x^{\omega}a_{6,K^{0}\omega}^c
+\frac{1}{2}(a_{7,K^{0}\omega}^c+a_{9,K^{0}\omega}^c) \nonumber\\
&-&\frac{1}{2}(a_{10,K^{0}\omega}^c
+\gamma_x^{\omega}a_{8,K^{0}\omega}^c)+\beta_3-\frac{1}{2}\beta_{3,EW}\Bigg]. \nonumber\\
\end{eqnarray}
and
\begin{eqnarray}
r'e^{i(\delta_{q}+\phi)}=\frac{p_{\omega}+p_{\omega}^{a}}{t_{\rho}+t_{\rho}^{a}}
=\frac{Q_4}{Q_2},
\end{eqnarray}

\begin{equation}
r'e^{i\delta_{q}}=\frac{Q_4}{Q_2}
\left|\frac{V_{tb}V_{td}^{*}}{V_{ub}V_{ud}^{*}}\right|,
\end{equation}
where
\begin{eqnarray}
\left|\frac{V_{tb}V_{td}^{*}}{V_{ub}V_{ud}^{*}}\right|=
\frac{\sqrt{[\rho(1-\rho)-\eta^{2}]^{2}+\eta^{2}}}{(1-\lambda^{2}/2)(\rho^{2}+\eta^{2})}.
\end{eqnarray}

It can be seen that $r'$ and $\delta_{q}$ depend on both the Wilson
coefficients and the CKM matrix elements, as shown in Eqs. (65).
Substituting Eqs. (58) (61) (65) into Eq. (49), we can obtain
$r$, $\sin\delta$, and $\cos\delta$. Then, in combination with Eqs.
(50) and (51) the $CP$ violating asymmetries can be obtained.

\section{Branching ratio of $\bar{B}_{s}^{0}\rightarrow K^{0}\rho^{0}(\omega)$}
\label{sec:11}
The matrix element for $B_s\rightarrow P$ and
$B_s\rightarrow
 V$ (where $P$ and $V$ denote pseudoscalar and vector mesons,
 respectively) can be decomposed as follows {\cite{Bauer1987}}:
\begin{eqnarray*}
\langle
P|J_{\mu}|B_s\rangle&=&\Bigg(p_{B_s}+p_{P}-\frac{m_{B_{s}}^{2}-m_{P}^{2}}{k^{2}}k\Bigg)
_{\mu}F_{1}(k^{2})   \nonumber \\
&+&\frac{m_{B_s}^{2}-m_{P}^{2}}{k^{2}}k_{\mu}F_{0}(k^{2}),
\end{eqnarray*}
\begin{eqnarray}
\langle
V|J_{\mu}|B_s\rangle&=&\frac{2}{m_{B_s}+m_{V}}\epsilon_{\mu\nu\rho\sigma}\epsilon^{*\nu}
p_{B_s}^{\rho}p_{V}^{\sigma}V(k^{2})   \nonumber \\
&+&i\Bigg\{\epsilon^{*}_{\mu}(m_{B_s}+m_{V})A_1({k^2})
-\frac{\epsilon^{*}\cdot k}{m_{B_s}+m_{V}}   \nonumber \\
&\times&(p_{B_s}+p_{V})_{\mu}A_2(k^2)-\frac{\epsilon^{*}\cdot
k}{k^2}2m_{V}\cdot k_\mu A_3(k^2)\Bigg\}  \nonumber
\\&+&i\frac{\epsilon^{*}\cdot k}{k^2}2m_{V}\cdot k_\mu A_0(k^2),
\end{eqnarray}
where $J_{\mu}$ is the weak current
($J_{\mu}=\bar{q}\gamma^{\mu}(1-\gamma_{5})b$ with $q=u,d,s$),
$p_{B_s} (m_{B_s}), p_{P} (m_{P}), p_{V} (m_{V})$ are the momenta
(masses) of $B_s, P, V$, respectively, $k=p_{B_s}-p_{P}(p_{V})$ for
$B_s\rightarrow P(V)$ transition and $\epsilon_{\mu}$ is the
polarization vector of $V$. $F_{i}$ $(i=0,1)$ and $A_{i}$
$(i=0,1,2,3)$ in Eq. (67) are the weak form factors which satisfy
$F_1(0)=F_0(0)$, $A_3(0)=A_0(0)$, and
$A_3(k^2)=[(m_B+m_{V})/2m_{V}]A_1(k^2)-[(m_B-m_{V})/2m_{V}]A_2(k^2)$.

 With the factorizable decay amplitudes in Eq. (56)(57) we can calculate the
decay rate for $B_s$ to a pseudoscalar meson $(P)$ and a vector meson
$(V)$ transition by using the following expression
{\cite{Hwang1999}}:
\begin{eqnarray}
\Gamma(B_s\rightarrow PV)&=&\frac{p_{c}}{8\pi m_{V}^{2}}
|A(B_s\rightarrow PV)/(\epsilon \cdot p_{B_s}) |^{2},
\end{eqnarray}
where
\begin{eqnarray*}
p_{c}=\frac{\sqrt{[m_{B_s}^{2}-(m_{P}+m_{V})^{2}][m_{B_s}^{2}-(m_{P}-m_{V})^{2}]}}{2m_{B_s}}
\end{eqnarray*}
is the c.m. momentum of the product particle and $A(B_s\rightarrow
PV)$ is the decay amplitude.

In the QCD factorization approach.
 Here $V_{u}^{T,P}$ are the CKM factors,
\begin{eqnarray}
V_{u}^{T}=|V_{ub}V_{uq}^{*}|, &        {\rm for} &  i=1,2,
\end{eqnarray}
and
\begin{eqnarray}
V_{u}^{P}=|V_{tb}V_{tq}^{*}|, &           {\rm for} & i=3,....,10.
\end{eqnarray}
In our case we take into account the $\rho-\omega$ mixing
contribution when we calculate the branching ratios since we are
working to the first order of isospin violation. we can
explicitly express the branching ratio for the process
$\bar{B}_{s}\rightarrow K^{0}\rho^{0}(\omega)$ as the following:
\begin{eqnarray}
&&BR(\bar{B}_{s}\rightarrow K^{0}\rho^{0}(\omega) ) \nonumber \\
&=&\frac{G_{F}^{2}p_{c}^3}{16\pi m_{\rho}^{2}\Gamma_{B_{s}}}\large|[V_u^TA_{\rho^0}^T(a_1,a_2)
-V_u^PA_{\rho^0}^P(a_3,...,a_{10})]    \nonumber  \\
&+&[V_u^TA_{\omega}^T(a_1,a_2)-V_u^PA_{\omega}^P(a_3,...,a_{10})] \nonumber  \\
&\times &\frac{\tilde{\Pi}_{\rho\omega}}
{(s_\rho-m_\omega^2)+im_\omega\Gamma_{\omega}}\large|^{2},
\end{eqnarray}
where $\Gamma_{B_{s}}$ is the total decay width of $B_{s}$.

\section{Input parameters}
\label{sec:7}In the numerical calculations, we have several
parameters, i.e. $N_c$ and the CKM matrix elements in the
Wolfenstein parametrization. For the CKM matrix elements, which should
be determined from experiments, we use the results of Ref.
\cite{Yao2006}:
\begin{eqnarray}
\bar{\rho}=0.132^{+0.022}_{-0.014},&&
\bar{\eta}=0.341\pm 0.013, \nonumber \\
\lambda=0.2253\pm0.0007, && A=0.808^{+0.022}_{-0.015}.
\end{eqnarray}
In QCD factorization scheme,
since power corrections have been considered,
$N_c$ is only color parameter, hence we use $N_c=3$. In naive factorization
$N_c$ includes the nonfatorizable
effects which are model and process dependent and cannot be
theoretically evaluated accurately and can be determined by
experiment.

The running quark masses is taken by the scale $\mu$ in $B_s$ decay.
One has
\begin{eqnarray}
 m_b(m_b)=4.2 GeV,  &&m_{c}(m_b)=0.91GeV,  \nonumber  \\
  m_u(m_b)=m_d(m_b)=0,  && m_s(2.1 GeV)=0.095 GeV.  \nonumber  \\
\end{eqnarray}

The values of the scale dependent quantities $f_V^\perp(\mu)$ and
$a_{1,2}^\perp(\mu)$ are given for $\mu=1GeV$. The value of Gegenbauer moments are
taken from {\cite{a2007}}.
\begin{eqnarray}
a_1^{\rho}=0,    &&a_2^{\rho} =0.15\pm 0.07   \nonumber \\
a_1^{\omega}=0,    &&a_2^{\omega} =0.15\pm 0.07   \nonumber \\
 a_{1}^{\perp\rho}=0, && a_{2}^{\perp\rho}=0.14\pm 0.06  \nonumber \\
  a_{1}^{\perp\omega}=0, && a_{2}^{\perp\omega}=0.14\pm 0.06 \nonumber \\
  a_1^{K}=0.06\pm 0.03,  &&a_2^{K}=0.25\pm 0.15  \nonumber \\
  f_{\rho}=216\pm 3 MeV,   &&f_{\rho}^\perp(\mu)=165\pm 9 MeV, \nonumber \\
f_{\omega}=187\pm 5 MeV,   &&f_{\omega}^\perp(\mu)=151\pm 9 MeV,
\end{eqnarray}

For $B_s$ meson, we use the value\cite{Yao2006}:
\begin{eqnarray}
 \tau=1.47 ps,  & m_{B_s}=5.366 GeV
 \end{eqnarray}
The Wilson coefficients $c_{i}$ can be found in {\cite{qcdf}}.
As discussed in detail in {\cite{qcdf}}, there are large theoretical
uncertainties related to the modeling of power corrections corresponding
to weak annihilation effects and the chirally-enhanced power corrections
to hard spectator scattering. So we parameterize
these effects in terms of the divergent integrals $X_H$ (hard spectator
scattering) and $X_A$ (weak annihilation). We also model these quantities by
using the parameterization{\cite{qcdf}}
\begin{equation}\label{XHparam}
   X_A = \left( 1 + \varrho_A\,e^{i\varphi_A} \right)
   \ln\frac{m_B}{\Lambda_h} \,; \qquad
   \varrho_A \le 1 \\
    \qquad \Lambda_h=0.5\,\mbox{GeV} \,,
\end{equation}
and similarly for $X_H$. Here $\varphi_A$ is an arbitrary
strong-interaction phase, which may be caused by soft rescattering. The
fitted $\varrho_A$ and $\varphi_A$ are taken from {\cite{chenga2009}}.
For $B_s\rightarrow PV$ decay, $\rho_A^{PV}\approx 0.87$, $\phi_A^{PV}\approx -30^\circ$.
For the estimate of theoretical uncertainties,
we shall assign an error of $\pm0.1$ to $\rho_A$ and $\pm 20^\circ$ to $\phi_A${\cite{chenga2009}}.

The form factors associated with the weak transitions depend on the
inner structure of the hadrons and are hence model dependent. Here
we will consider the form factors obtained in several
phenomenological models.  For $B_{s}$ decay form
factors, we will use the results (form factors
are referred to the ones at $q^2=0$):
\renewcommand{\theenumi}{\arabic{enumi})}
\begin{enumerate}
\item
Model 1 {\cite{qcdf}}\\
\begin{eqnarray*}
F_{0}^{B_s\rightarrow K}=0.31\pm0.05,
\end{eqnarray*}
\item
Model 2 (in the pQCD approach)\cite{AliBs}\\
\begin{eqnarray*}
F_0^{B_s\rightarrow K}=0.24^{+0.05+0.00}_{-0.04-0.01},
\end{eqnarray*}
\item
Model 3 (form factors obtained by QCD sum rules)\cite{Melic}\\
\begin{eqnarray*}
F_0^{B_s\rightarrow K}=0.30^{+0.04}_{-0.03},
\end{eqnarray*}
\item
Model 4
(light-cone sum rule calculation based on heavy quark effective theory)\cite{WuFF}\\
\begin{eqnarray*}
F_0^{B_s\rightarrow K}=0.296\pm0.018,
\end{eqnarray*}
\item
Model 5
(A light cone quark model in conjunction with soft collinear effective theory)\cite{LuFF}\\
\begin{eqnarray*}
F_0^{B_s \rightarrow K}=0.290,
\end{eqnarray*}
\item
Model 6
(lattice QCD calculation)\cite{lattice}\\
\begin{eqnarray*}
 F_0^{B_s \rightarrow K}=0.23\pm0.05\pm0.04.
 \end{eqnarray*}
\end{enumerate}

In above Models, the $k^{2}$ dependence of the form factors has the
following form under the nearest pole dominance assumption:
\begin{eqnarray}
h(k^{2})=\frac{h(0)}{1-\frac{k^{2}}{m_{h}^{2}}},
\end{eqnarray}
where $h$ could be $F_{0}$, and
$m_{h}$ is the pole mass.

It is noted that since the value of $k^2$ (which is actually the
square of the mass of the factorized light meson) is much smaller
than the square of the pole mass which is of order $m_b^2$, only the
values of the form factors at $k^2=0$ are most relevant and hence
how the form factors depend on $k^2$ has little effects (less than
2\%). From the above values we see that the form factor $B_s\rightarrow K$
at $q^2=0$ ranges from $0.23$ to $0.31$.

\section{Numerical results and discussions}
\subsection{$CP$ violation via $\rho-\omega$ mixing in $\bar{B}_{s}^{0}\rightarrow K^{0}\pi^{+}\pi^{-}$}
\label{sec:8}In the numerical calculations, we find the $CP$ violating
asymmetry, $a$, is large when the
invariant mass of $\pi^{+}\pi^{-}$ is in the vicinity of the
$\omega$ resonance within QCD factorization scheme.

In the respective error ranges, when $\sqrt{s}=0.782$ $GeV$, we get
maximum $CP$ violating asymmetry
\begin{eqnarray}\label{a value}
a=(45.9^{+16.2+27.5}_{-15.7-26.1})\times 10^{-2}
\end{eqnarray}

\begin{figure}
\resizebox{0.5\textwidth}{!}{%
  \includegraphics{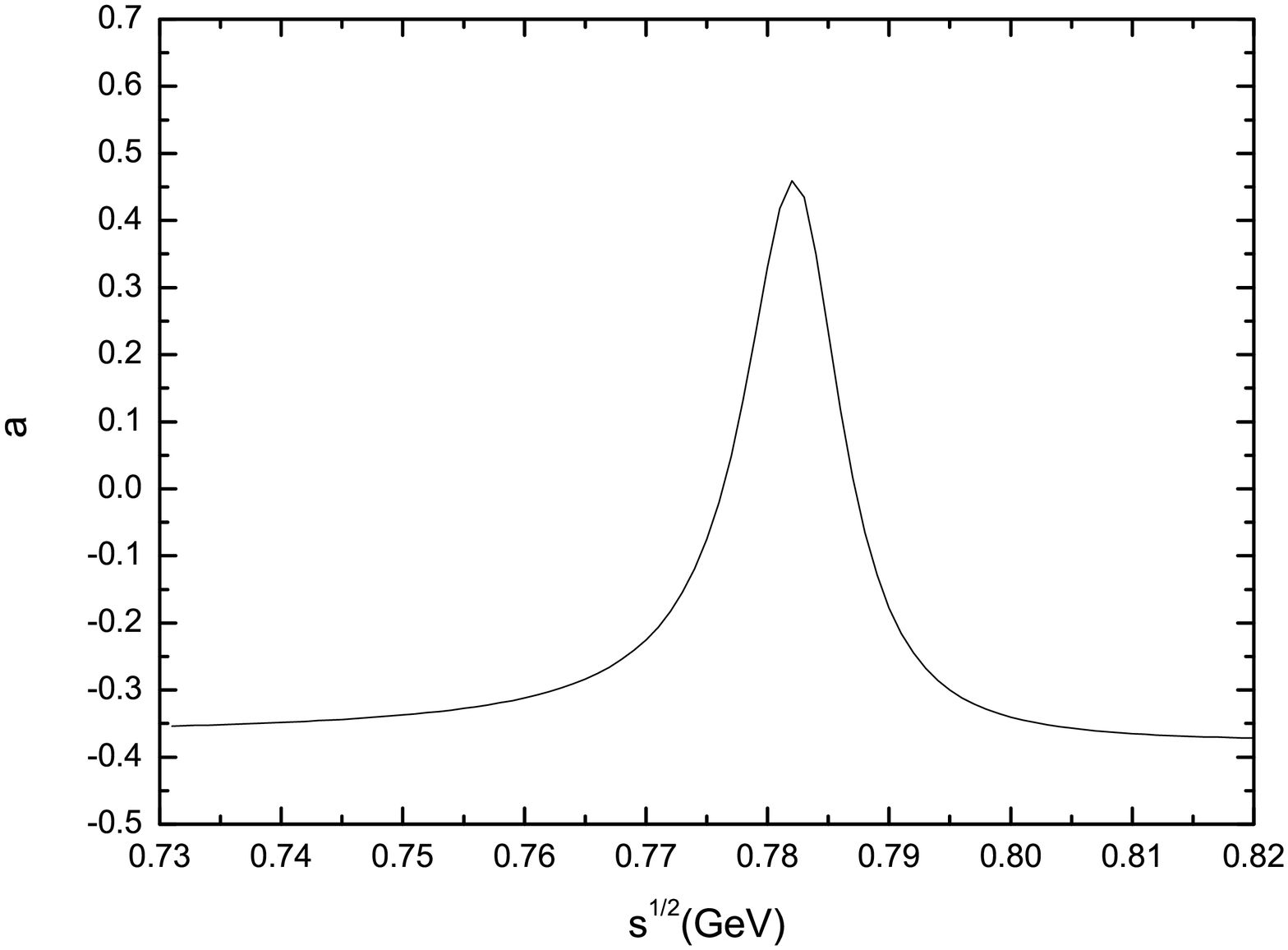}
  }
\caption{Plot of $a$ as a function of $\sqrt{s}$
corresponding to central parameter values of CKM matrix elements
for $\bar{B}_{s}^{0}\rightarrow K^{0}\rho^{0}(\omega)\rightarrow K^{0}\pi^{+}\pi^{-}$.
}
\label{fig:1}       
\end{figure}

\begin{figure}
\resizebox{0.5\textwidth}{!}{%
  \includegraphics{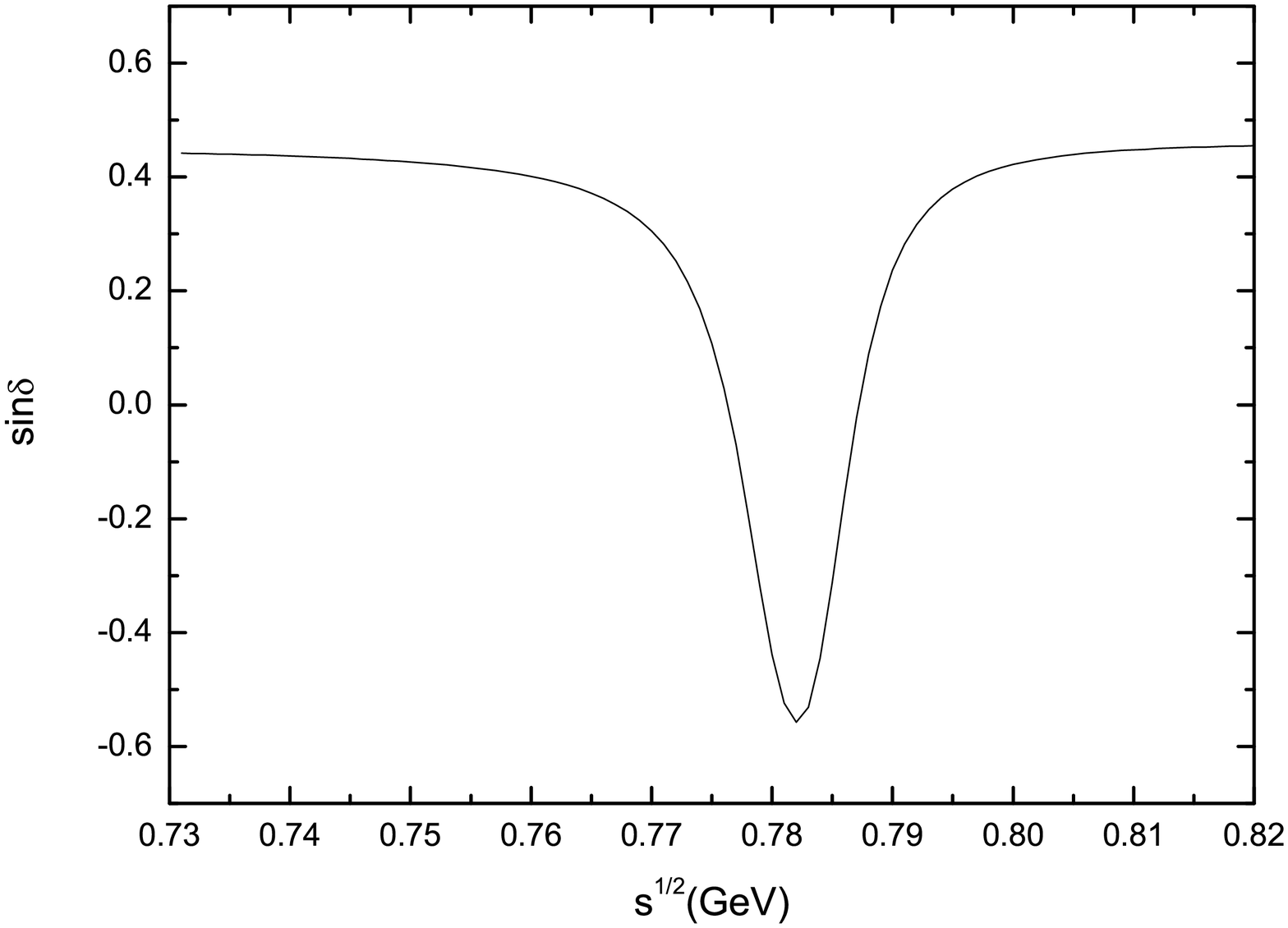}

}
\caption{Plot of $\sin\delta$ as a function of $\sqrt{s}$
corresponding to central parameter values of CKM matrix elements
 with $\rho-\omega$ mixing for $\bar{B}_{s}^{0}\rightarrow K^{0}\rho^{0}(\omega)
 \rightarrow K^{0}\pi^{+}\pi^{-}$.}
\label{fig:2}       
\end{figure}

\begin{figure}
\resizebox{0.5\textwidth}{!}{%
  \includegraphics{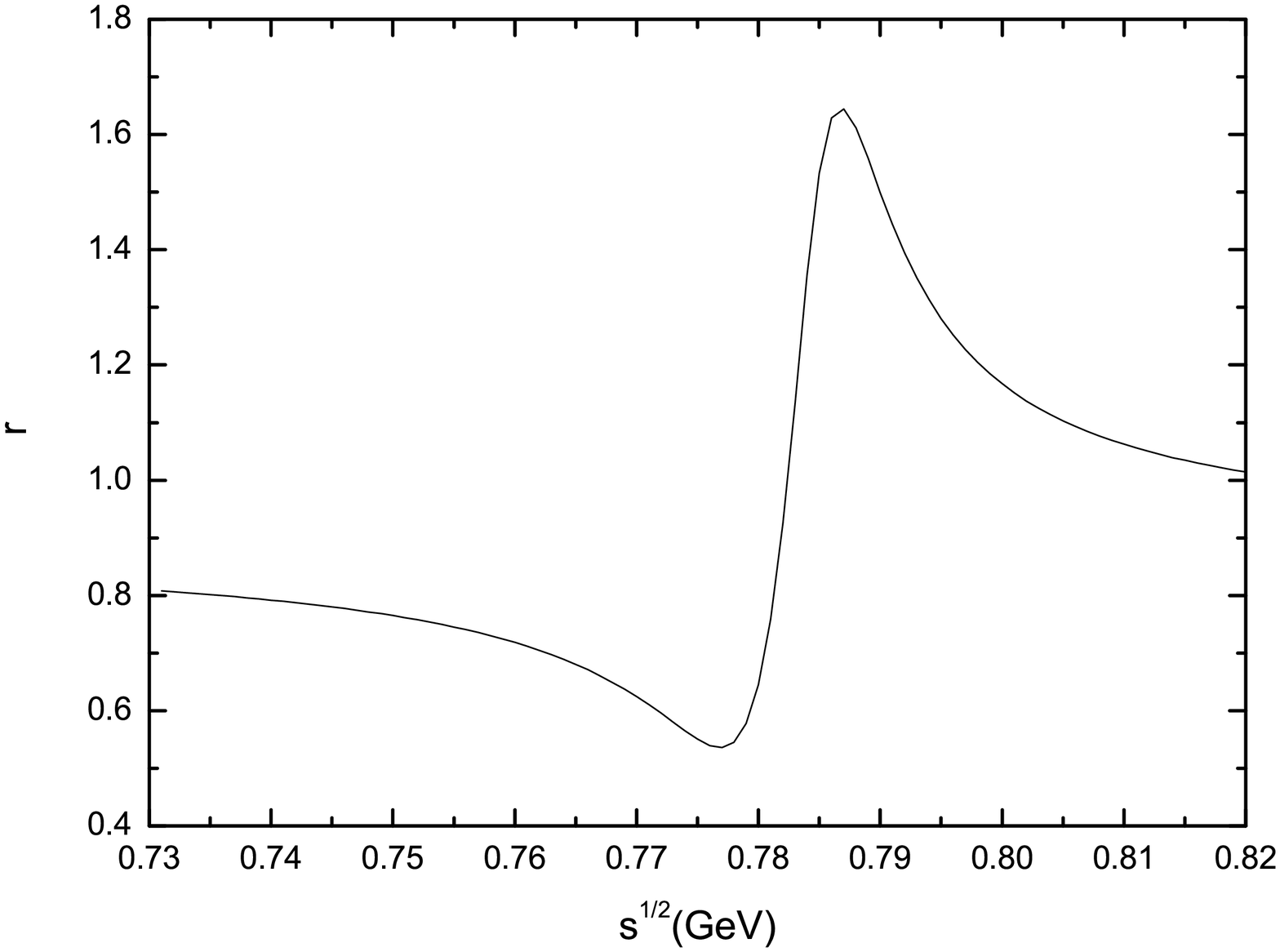}

}
\caption{Plot of $r$ as a function of $\sqrt{s}$
corresponding to central parameter values of CKM matrix elements
 with $\rho-\omega$ mixing for $\bar{B}_{s}^{0}\rightarrow K^{0}\rho^{0}(\omega)
 \rightarrow K^{0}\pi^{+}\pi^{-}$.}
\label{fig:3}       
\end{figure}

In QCD factorization, the theoretical errors are large which follows to
the uncertainties of results. Generally, power corrections beyond the heavy
quark limit give the major theoretical uncertainties. This implies the necessity of introducing $1/m_b$ power corrections. Unfortunately, there are many possible $1/m_b$ power suppressed effects and they are generally nonperturbative in nature and hence not calculable by the perturbative method.
There are more uncertainties in this scheme.
The first error refers to the variation of the CKM parameters. The second error
comes from form factors and decay constants. The third error corresponds to
the Gegenbauer moments. The last error is the wave function of the $B_s$ meson characterized by the
parameter $\lambda_B$, the power corrections due to weak annihilation and hard
spectator interactions described by the parameters $\rho_{A,H}$, $\phi_{A,H}$,
respectively. Using the central values of above parameters,
we first calculate the numerical results of $CP$ violation and branching ratio,
and then add errors according to standard deviation. In Fig.1, We
give the central value of $CP$ violating asymmetry
as a function of $\sqrt{s}$. From the figure one can see
the CP asymmetry parameter is dependent on $\sqrt{s}$
and changes rapidly due to $\rho-\omega$ mixing when the
invariant mass of $\pi^{+}\pi^{-}$ is in the vicinity of the
$\omega$ resonance. The CP violating asymmetry vary
from around $-37\%$ to around $45\%$.

From Eq. (43), one can see that the $CP$ violating  asymmetry
parameter depends on both $\sin\delta$ and $r$. The plots of
$\sin\delta$ and $r$ as a function of $\sqrt{s}$
 are shown in Fig. 2 and Fig. 3.  It
can be seen that when $\rho-\omega$ mixing is taken into account
 $\sin\delta$ and $r$ change sharply when the invariant mass of
$\pi^{+}\pi^{-}$ is around 0.782 GeV. From the Fig. 2, one can
find $\rho-\omega$ mixing make the $\sin\delta$ value oscillate from
$-0.56$ to $0.44$ which can not reach the value $-1$. This result
is not in agreement with conclusion from naive factorization which can
measure the $CP$ violating parameter to remove the
mod($\pi$) phase uncertainty in the determination of the CKM angle
$\alpha$ arising from the conventional determination through
$\sin2\alpha ${\cite{Leitner2003}}.

We have shown that $\rho-\omega$ mixing does enhance the direct $CP$
violating asymmetries and provide a mechanism for large $CP$
violation in QCD factorization scheme. On the other hand, it
is important to see whether it is possible to
observe these large $CP$ violating asymmetries in experiments. This
depends on the branching ratio for $\bar{B}_{s}^{0}\rightarrow K^{0}\rho^{0}(\omega)$.
We will study this problem in the next section.

\subsection{Branching ratios via $\rho-\omega$ mixing in $\bar{B}_{s}^{0}\rightarrow K^{0}\rho^{0}(\omega)$}

Including $\rho-\omega$ mixing, we calculate the values
of branching ratios for $\bar{B}_{s}^{0}\rightarrow K^{0}\rho^{0}(\omega)$.
Base on the reasonable parameters range, we obtain the branching ratio of
$\bar{B}_{s}^{0}\rightarrow
K^{0}\rho^{0}(\omega)$ is $(9.8^{+2.6+3.4}_{-1.2-0.7})\times 10^{-7}$ which
is consistent with the result {\cite{chenga2009}}. In other words,
although we calculate the branching ratio due to
 $\rho-\omega$ mixing in QCD factorization scheme,
 we find the contribution  of  $\rho-\omega$ mixing
 for branching ratio is small and can be neglected.
 $\rho-\omega$ mixing mechanism presents new phase differences and
 produce extremely small effect for branching ratio of $\bar{B}_{s}^{0}\rightarrow K^{0}\rho^{0}(\omega)$.

\section{Discussions on possibility to observe $CP$ violating asymmetries at the LHC}
\label{sec:16}The LHC is a proton-proton collider currently have started
at CERN. With the designed
center-of-mass energy $14$ TeV and luminosity $L=10^{34}
cm^{-2}s^{-1}$, the LHC gives access to high energy frontier at TeV
scale and an opportunity to further improve the consistency test for
the CKM matrix. The production rates for heavy quark flavours will
be large at the LHC, and the $b\bar{b}$ production cross section
will be of the order 0.5 mb, providing as many as $0.5\times
10^{12}$ bottom events per year {\cite{Schopper2005}}. In
particular, the LHCb detector is designed to exploit large number of
$b$-hadrons produced at the LHC in order to make precise studies on
$CP$ asymmetries and on rare decays in $b$-hadron systems. The other
two experiments, ATLAS and CMS, are optimized for discovering new
physics and will complete most of their $B$ physics program within
the first few years {\cite{Schopper2005,Gouz2004}}. Obviously, the
LHC has a great advantage over the current experiments on
$b$-hadrons{\cite{Belle2002}}.

In the present work, we have predicted possible large $CP$ violating
asymmetries in decay channel of $\bar{B}_{s}^{0}\rightarrow K^{0}\rho^{0}(\omega)
\rightarrow K^{0}\pi^{+}\pi^{-}$ via the $\rho-\omega$ mixing.
At the LHC, the $b$-hadrons are produced
from the $pp$ collisions. The possible asymmetry between the numbers
of the $b$-hadrons, $H_b$, and those of their antiparticles,
$\bar{H_{b}}$, has been studied in the Lund string fragmentation
model and the intrinsic heavy quark model {\cite{lhc1,lhc2}}. It has
been shown that this asymmetry can only reach values of a few
percent. In our following discussions, we will ignore this small
asymmetry and give the numbers of $H_{b}\bar{H_{b}}$ pairs needed
for observing the $CP$ violating asymmetries we have predicted.
These numbers depend on both the magnitudes of the $CP$ violating
asymmetries and the branching ratios of heavy hadron decays which
are model dependent. For one (three) standard deviation signature,
the number of $H_{b}\bar{H_{b}}$ pairs we need is
{\cite{Du1986,Louis,WT}}
\begin{eqnarray}
N_{H_{b}\bar{H_{b}}}\sim \frac{1}{BRa^{2}}(1-a^{2})\Bigg(\frac{9}{BR
a^{2}}(1-a^{2})\Bigg),
\end{eqnarray}
where BR is the branching ratio for $H_{b}\rightarrow f\rho^{0}$.

For central value of $CP$ asymmetry in Eq. ({\ref{a value}}), we
present the numbers of $B_{s}\bar{B}_{s}$ pairs for observing
the large $CP$ violating asymmetries at LHC.
For the channel
 $\bar{B}_{s}^{0}\rightarrow K^{0}\rho^{0}(\omega)\rightarrow K^{0}\pi^{+}\pi^{-}$ ,
the numbers of  $B_{s}\bar{B}_{s}$ pairs are $3.8\times 10^{6}$ ($3.4\times10^{7}$)
for $1\sigma$ ($3\sigma$) signature.
At the LHC the average
$B_{s}\bar{B_{s}}$ production is about $10\% $ out of $10^{12}$
$b\bar{b}$ events {\cite{Schopper2005}}.
From Fig.1, one can see $CP$ violating asymmetries vary sharply at small energy range,
and reach peak value at $\sqrt{s}=0.782$ $GeV$. Hence, it is very
possible to observe the large $CP$ violating asymmetries in small
energy range of $\rho^{0}\sim \omega$ resonance at the peak values of $CP$ violating asymmetries
from the LHC experiment. For the experiments, it is possible to reconstruction $\pi^{+}$, $\pi^{-}$ and $K^0$ mesons
when the invariant
masses of $\pi^{+}\pi^{-}$ pairs are in the vicinity of the $\omega$
resonance. Therefore, it is very
possible to observe the large $CP$ violating asymmetries in
$\bar{B}_{s}^{0}\rightarrow K^{0}\rho^{0}(\omega)\rightarrow
K^{0}\pi^{+}\pi^{-}$ at the LHC.
\\

\section {Summary and conclusions}
\label{sec:17} In this paper, we have studied $CP$ violation in
$\bar{B}_{s}^{0}\rightarrow K^{0}\rho^{0}(\omega)\rightarrow
K^{0}\pi^{+}\pi^{-}$. It has been found that, by including $\rho-\omega$ mixing, the $CP$
violating asymmetries can be large when the invariant
masses of $\pi^{+}\pi^{-}$ pairs are in the vicinity of the $\omega$
resonance. For the decay $\bar{B}_{s}^{0}\rightarrow K^{0}\rho^{0}(\omega)\rightarrow
K^{0}\pi^{+}\pi^{-}$, the maximum CP violation can reach $46\%$. Furthermore, taking
$\rho-\omega$ mixing into account, we have calculated the branching
ratio of the decay. We have also presented
the numbers of $B_{s}\bar{B_{s}}$ pairs required for observing the
predicted $CP$ violation in experiments at the LHC. We have found
the channel is the likely channel in which the large $CP$ violating
asymmetries may be observed at LHC. We expect that our
predictions will provide a useful guidance for future investigations
and experiments.

In our calculations there are some uncertainties. We have worked in
the QCD factorization which is expected to be a reliable
approach in the heavy-quark limit. In the QCD factorization scheme,
$\alpha_{s}(m_{b})$ and some $1/m_b$ (annihilation) corrections are
included. In this framework, there is cancellation of the scale and
renormalization scheme dependence between the Wilson coefficients
and the hadronic matrix elements. However,
the QCD factorization scheme suffers from endpoint
singularities which are not well controlled. The $CP$ violating
asymmetry depends on the unknown parameters which are associated
with such endpoint singularities. The CKM matrix elements also lead to some
uncertainties in the $CP$ violating asymmetry. Uncertainties also
come from the weak form factors associated with the hadronic matrix
elements. This lead to uncertain $CP$
violating asymmetries in the QCD factorization scheme. This needs
further detailed investigations.

{\it Acknowledgements.} This work was supported by
 the Special Grants (Project Number 2009BS028) for PH.D
 from Henan University of Technology.

%

\end{document}